\documentclass[aps,prd,onecolumn,notitlepage,preprint, showkeys,superscriptaddress]{revtex4-2}

\usepackage{graphicx}
\usepackage{epstopdf}
\usepackage{amsmath}
\usepackage{amssymb}
\usepackage[section]{placeins}
\usepackage{slashed}
\usepackage{color}
\usepackage{soul}
\usepackage{appendix}
\usepackage{placeins}
\usepackage[utf8]{inputenc}
\usepackage{pstricks}
\usepackage{multirow}
\usepackage{hyperref}
\usepackage{subfigure}
\usepackage{epsfig}
\usepackage{cancel}
\usepackage{braket}
\usepackage[normalem]{ulem} 
\newcommand{\be}{\begin{equation}}
\newcommand{\ee}{\end{equation}}
\newcommand{\ben}{\begin{eqnarray}}
\newcommand{\een}{\end{eqnarray}}

\usepackage{subcaption}
\usepackage{dcolumn}
\usepackage{bm}
\usepackage{physics}
\usepackage{lipsum}
\usepackage{tikz-feynman}
\usepackage[normalem]{ulem}   
\usepackage{xcolor}           
\usepackage{hyperref}         

\newcommand{\UFBA}{Instituto de Física, Universidade Federal da Bahia, Campus Ondina, Salvador, Bahia 40170-115, Brazil}

\newcommand{\Pedro}[1]{\textcolor{blue}{#1}}
\begin{document}

\title{Searching for the pseudoscalar partner of $G(3900)$ via radiative $Y(4230)$ decays}

\author{Vitor C. Premoli}
\email{vitorpremoli95@gmail.com}
\affiliation{\UFBA}

\author{Pedro Brandão}
\email[]{pedro.brandao@ufba.br}
\affiliation{\UFBA}

\author{Ya-Wen Pan}
\email[]{pyw1519@buaa.edu.cn}
\affiliation{School of Physics, Beihang University, Beijing 102206, China}

\author{Li-Sheng Geng}
\email[]{lisheng.geng@buaa.edu.cn}
\affiliation{Sino-French Carbon Neutrality Research Center, \'Ecole Centrale de P\'ekin/School of General Engineering, Beihang University, Beijing 100191, China}
\affiliation{School of Physics, Beihang University, Beijing 102206, China}
\affiliation{Peng Huanwu Collaborative Center for Research and Education, Beihang University, Beijing 100191, China}
\affiliation{Southern Center for Nuclear-Science Theory (SCNT), Institute of Modern Physics, Chinese Academy of Sciences, Huizhou 516000, China}

\author{Luciano M. Abreu}
\email[]{luciano.abreu@ufba.br}
\affiliation{\UFBA}


\begin{abstract}

Inspired by the $P$-wave molecular interpretation of the recently observed vector state $G(3900)$, we analyze the production of its possible pseudoscalar partner, denoted here as $G_0(3900)$, via the radiative decay $Y(4230) \to \gamma G_0(3900)$. The $G_0(3900)$ is interpreted as a $P$-wave molecular state with quantum numbers $J^{PC}=0^{-+}$, dominated by the $D\bar{D}^{\ast}/\bar{D}D^{\ast}$ components. Although not yet experimentally established, such a structure is expected to appear near the $D\bar{D}^*$ threshold and to exhibit characteristic production patterns. The decay is assumed to proceed through a triangle mechanism. Depending on the model parameters and the binding energy of  $G_0(3900)$, the resulting branching fraction lies in the range $\mathcal{B}(Y(4230) \to \gamma G_0(3900)) \approx 1 \times  10^{-5} - 8 \times  10^{-5}$.  Our results offer a pathway to search for signatures of $G_0(3900)$ in radiative channels and also provide a test of the consistency of loop-mediated radiative decays with a molecular description of the $Y(4230)$.

\end{abstract}

\maketitle
\newpage


\section{Introduction}

Over the past two decades, hadron spectroscopy has undergone a dramatic expansion. Spearheaded by the discovery of the $X(3872)$ in 2003 by the Belle Collaboration~\cite{Belle:2003nnu}, a rich spectrum of so-called ``exotic'' states has emerged---particularly in the charmonium and bottomonium regions---that cannot be adequately described by the simple quark–antiquark picture. These states, often labelled as the $X$, $Y$, and $Z$ heavy quarkonia, have stimulated a variety of theoretical interpretations, including compact multiquarks, hybrid states, glueballs, and hadronic molecules~\cite{ParticleDataGroup:2024cfk,Guo:2017jvc,Liu:2024uxn,Wang:2025dur}.

Among these, we remark the structure denoted as $G(3900)$ with quantum numbers $J^{PC} = 1^{--}$, first reported by the BaBar Collaboration in 2007~\cite{BaBar:2006qlj} and later confirmed by Belle in 2008~\cite{Pakhlov:2008cq} through analyses of the $e^+e^- \to D\bar{D}$ cross section. However, at that time the enhancement was not considered evidence for a genuine resonance, as it could be explained by alternative means, such as kinematic effects~\cite{Huang:2025rvj}. The situation changed recently when the BESIII Collaboration reported a structure in the same energy region with high significance~\cite{BESIII:2024ths}, having a mass of $(3872.5 \pm 14.2 \pm 3.0)~\mathrm{MeV}$ and a width of $(179.7 \pm 14.1 \pm 7.0)~\mathrm{MeV}$.
This confirmation has prompted renewed interest, and the $G(3900)$ has recently been proposed to be a $P$-wave molecular state, predominantly composed of $D\bar{D}^*/D^*\bar{D}$ mesons~\cite{Chen:2025gxe,Lin:2024qcq, Ye:2025ywy}.

The interpretation of the $G(3900)$ as a $P$-wave molecule has, in turn, stimulated theoretical exploration of other $P$-wave molecular configurations near the $D\bar{D}^{\ast}/\bar{D}D^{\ast}$ threshold. In particular, Refs.~\cite{Chen:2025gxe,Lin:2024qcq,Liu:2025sjz} have systematically examined the possibility of a pseudoscalar partner --- denoted here as $G_0(3900)$ --- with quantum numbers $I^G(J^{PC}) = 0^+(0^{-+})$, whose structure is dominated by the $D\bar{D}^{\ast}/\bar{D}D^{\ast}$ channels, expressed as:
\begin{equation}
|G_0\rangle = \frac{1}{\sqrt{2}} \bigl( |D\bar{D}^*\rangle - |\bar{D}D^*\rangle \bigr),
\label{G_0}
\end{equation}
with
\begin{equation}
|D\bar{D}^*\rangle = \frac{1}{\sqrt{2}} \bigl( |D^0\bar{D}^{*0}\rangle + |D^+\bar{D}^{*-}\rangle \bigr).
\label{G_0bis}
\end{equation}
Both Refs.~\cite{Chen:2025gxe,Lin:2024qcq} employed models based on the one-boson-exchange interaction. Ref.~\cite{Lin:2024qcq} predicts a resonant pole associated with this state, which moves to the physical Riemann sheet and becomes a bound state as the cutoff is increased to strengthen the attraction; in this bound-state configuration, the binding energy relative to the $D\bar{D}^{\ast}/\bar{D}D^{\ast}$ threshold varies from 0 to 40~MeV. Meanwhile, Ref.~\cite{Liu:2025sjz} finds that the state may manifest itself as a resonance, a virtual state or a bound state with binding energy up to 50 MeV depending on the interaction strength.

While the $G_0(3900)$ state has yet to be experimentally established, Ref.~\cite{Liu:2025sjz} provides concrete predictions for its hadronic decay patterns within a molecular framework, calculating the partial decay widths of $G_0$ into $\omega(\rho^0) J/\psi$, $\pi^+\pi^- \eta_c(1S)$, $\pi^+\pi^- \chi_{c1}(1P)$, and $D^0\bar{D}^0\pi^0$. Varying the model parameters and assuming a binding energy between 0.1 and 10~MeV, the authors find that the hidden-charm decays are dominated by $G_0 \to \omega J/\psi$ and $G_0 \to \pi^+\pi^-\eta_c(1S)$, with partial widths reaching up to $1$~MeV and $0.1$~MeV, respectively. Whereas, the decay width to the open-charm channel $G_0 \to D^0\bar{D}^0\pi^0$ is predicted to be much smaller, at the order of approximately $0.1$~keV. Based on these findings, the authors suggest that BESIII and Belle II should search for the $G_0(3900)$ in the hidden-charm processes $G_0 \to \omega J/\psi$ or $G_0 \to \pi^+\pi^-\eta_c(1S)$. 


The present work aims to provide a complementary probe of the $G_0(3900)$ state by studying its production in a different reaction. We investigate the radiative decay
$Y(4230) \to \gamma \ G_0(3900)$,
where $Y(4230)$ ($J^{PC}=1^{--}$) is a well-established vector charmonium-like state. While the $Y(4230)$ has been subject to various interpretations in the literature~\cite{Ding:2008gr,MartinezTorres:2009xb,Wang:2013cya,Cleven:2013mka,Li:2013ssa,Wang:2013kra,Maiani:2014aja,Chen:2019mgp}, its nearness  to the $\bar{D}D_1$ threshold has motivated its description as a $S$-wave $D_1\bar{D}$ molecule~\cite{Ding:2008gr}. 
This proximity enables a triangle diagram production of the $D\bar{D}^*$-molecular candidate $G_0(3900)$: the $Y(4230)$ decays into $D_1\bar{D}$, the $D_1$ radiatively transitions to $D^*\gamma$, and the $D^*\bar{D}$ pair rescatters into $G_0$. This radiative decay chain thus offers both a plausible production mechanism and a sensitive diagnostic of the molecular nature of the participating states, as explored in other cases~\cite{Britto:2024nrp,Liu:2024ziu}.
In particular, according to the conventional meson picture, the decay $Y(4230) \to \gamma G_0(3900)$ involves a $1^{--} \to 0^{-+}$ radiative transition, which, by angular momentum and parity conservation, corresponds to an $M1$ transition at the level of the overall process. This should be distinguished from the internal $D_1 \to D^* \gamma$ transition, which is an $E1$ transition. The interplay between the overall $M1$ character of the $Y(4230) \to \gamma G_0(3900)$ process and the internal $E1$ nature of the $D_1 \to D^* \gamma$ subprocess is a distinctive feature of the triangle mechanism considered here. 
Quantifying the partial width for this transition therefore addresses two connected goals: assessing the observability of $G_0(3900)$ in radiative channels, and testing whether loop-mediated radiative decay amplitudes are consistent with the molecular picture of $Y(4230)$.
To obtain an estimate of the partial decay width, we employ an effective Lagrangian framework that captures the relevant vertices and evaluate the triangular loop diagrams in the isospin limit. We explore the sensitivity of the predicted width to two key parameters: the binding energy $E_b$, which parametrizes the mass of the unobserved $G_0(3900)$, and the size parameter $\Lambda$ of the form factor, which encodes short-distance physics not explicitly included in the effective theory.


This paper is organized as follows. Section~\ref{Formalism} presents the effective Lagrangians, derives the loop amplitude, and discusses the tensor reduction and pole structure used to perform the integrals. Section~\ref{results} contains numerical results and sensitivity studies for the size parameter $\Lambda$, binding energy $E_b$, and the effect of a finite $D_1$ width. Concluding remarks are given in Section~\ref{Conclusions}. 


\section{Formalism}\label{Formalism}


\begin{figure}[!htbp] 
\centering
\includegraphics[width=0.8\linewidth]{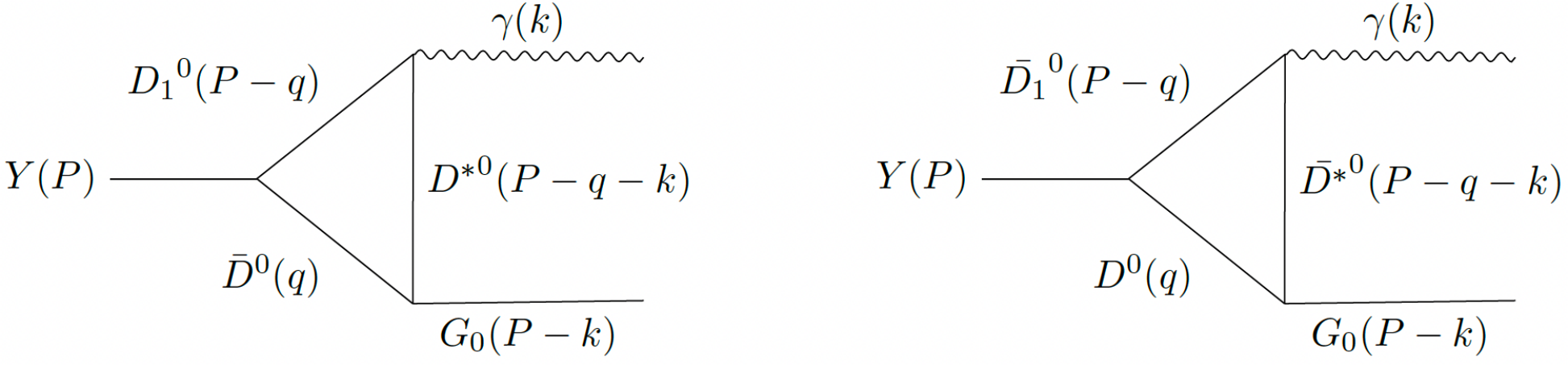}
\caption{Triangle diagrams contributing to the decay $Y(4230) \to \gamma G_0(3900)$ involving the neutral components of $G_0(3900)$. The respective momenta of the particles are in parentheses. \label{fig:neutral_triangle}}
\end{figure} 

\begin{figure}[!htbp]
\centering
\includegraphics[width=0.8\linewidth]{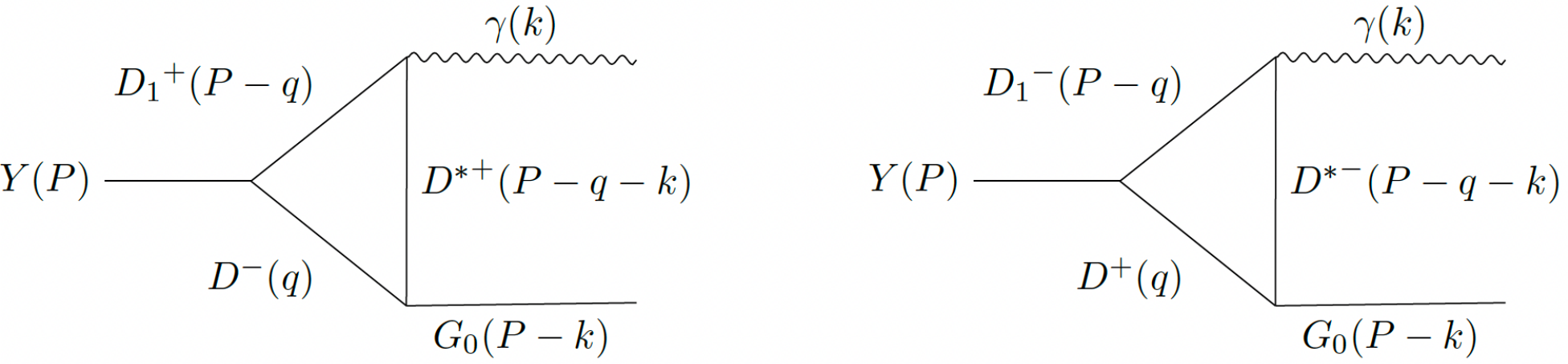}
\caption{Triangle diagrams contributing to the decay $Y(4230) \to \gamma \ G_0(3900)$ involving the charged components of $G_0(3900)$. The respective momenta of the particles are in parentheses. \label{fig:charged_triangle}}
\end{figure}

The radiative decay $Y(4230) \to \gamma G_0(3900)$ is analyzed here via a triangle loop mechanism, which proceeds as follows. The $Y(4230)$ first decays into $D_1\bar{D}$; the $D_1$ then undergoes a radiative transition to $D^{\ast}\gamma$; finally, the $D^{\ast}\bar{D}$ pair rescatters to form the $G_0(3900)$. Assuming that $G_0(3900)$ is a $P$-wave molecular state dominated by the $D\bar{D}^{\ast}/\bar{D}D^{\ast}$ channels, with its neutral and charged components given in Eq.~\eqref{G_0}, the corresponding triangle diagrams are shown in Figs.~\ref{fig:neutral_triangle} and~\ref{fig:charged_triangle}.
To calculate the respective amplitudes, we employ the effective Lagrangian approach. Since our main goal is to provide a first exploratory analysis of $G_0(3900)$ production via radiative decays, we simplify the calculation by working in the isospin limit. Within this approximation, the members of each isospin doublet become degenerate in mass, and as a consequence, all diagrams contribute equally to the transition amplitude. 

Accordingly, the effective Lagrangians for the vertices $Y^\mu \to D_1^\mu + \bar{D}$, $D_1 \to D^* + \gamma$ and $D^* + \bar{D} \to G_0$ are, respectively, 
\begin{align}
        \mathcal{L}_{YD_1D} & = g_{YD_1D}\bar{D}D_1^\mu Y_\mu + h.c , \nonumber \\ 
    \mathcal{L}_{D_1D^*\gamma} & = eg_{D_1D^*\gamma} \epsilon_{\mu\nu\alpha\beta}\left(\partial_\mu A_\nu D_{1\alpha} {D}^*_\beta\right) + h.c, \nonumber \\ 
    \mathcal{L}_{G_0DD^*} & = g_{G_0DD^*} G_0 {[D({\partial^\mu}\bar{D}^*_\mu) - ({\partial^\mu}D)\bar{D}^*_\mu]}  +h.c ,
    \label{Lagr1}
\end{align}
where the coupling constants have been fixed as follows. The coupling  $g_{YD_1\bar{D}}$ is determined to be $g_{YD_1\bar{D}} = 18.29 \times 10^{3}$~MeV~\cite{Liu:2024ziu}. 
Since we work in the isospin limit, we use the relations between this coupling in the particle basis to those in the isospin basis according to Ref.~\cite{Liu:2024uxn}. 
For the coupling $g_{D_1D^*\gamma}$, we use $g_{D_1D^{*}\gamma} = (g_{D_1^0 D^{*0}\gamma} + g_{D_1^+ D^{*+}\gamma})/2$, with the individual values $g_{D_1^0 D^{*0}\gamma} = 1.82$ and $g_{D_1^+ D^{*+}\gamma} = 0.58$, also obtained from Ref.~\cite{Liu:2024ziu}. 
The coupling constant $g_{G_0DD^*}$ is determined via the compositeness condition~\cite{Weinberg:1965zz,Gamermann:2009uq,Chen:2017abq,Sakai:2020ucu}. In the specific case considered here, we follow the procedure established in Ref.~\cite{Liu:2025sjz}, which also treats the $G_0(3900)$  as a $P$-wave $D\bar{D}^{*}/\bar{D}D^{*}$ molecular state. For completeness, we outline the main steps of this determination in Appendix~\ref{appa}, while referring the reader to Ref.~\cite{Liu:2025sjz} for further details.


Making use of the effective Lagrangians defined above, the amplitude for the processes illustrated in Figs.~\ref{fig:neutral_triangle} and ~\ref{fig:charged_triangle} can be written in the following compact form: 
\begin{eqnarray}
       \mathcal{M} & = & g_{YD_1D} \ eg_{D_1D^*\gamma} \ g_{G_0DD^*}\int \frac{d^4q}{(2\pi)^4}
        \frac{\left(-g^{\mu\nu} + \frac{(P - q)^\mu (P - q)^\nu}{m^2_{D_1}}\right)}{\big[q^2 - m_D^2\big]\big[(P - q)^2 - m_{D_1}^2\big]\big[(P - q - k)^2 - m_{D^*}^2\big]} \nonumber \\ 
        & & \times \left(- (P - 2q - k)^\delta + (P - q - k)^\delta \frac{(P - q - k)^2 - q \cdot (P - q - k)}{m^2_{D^*}}\right) \nonumber \\ 
        & & \times\epsilon_{\alpha\beta\nu\delta}k^\alpha \varepsilon^{*\beta}_{\gamma}(k)\epsilon_{Y\mu}(P) , 
\label{ampl1}
\end{eqnarray}
where $\epsilon_{Y} $ and $\epsilon_\gamma $ are the $Y(4230)$ and photon polarization vectors, respectively. 

The amplitude in Eq.~(\ref{ampl1}) is simplified by applying the transversality and on-shell conditions for the external particles.  
To employ the Passarino–Veltman method~\cite{Passarino:1978jh}, it is convenient to rewrite it as 
\begin{equation}
            \mathcal{M} = \left( \mathcal{M}_1 P^\delta + \mathcal{M}_2^\delta  \right) \ \epsilon_{\alpha \beta \nu \delta}  k^\alpha \varepsilon^{*\beta}_\gamma(k)\varepsilon^\nu_Y(P) + \mathcal{M}_3^{\nu \mu} \ \epsilon_{\alpha \beta \nu \delta} k^\alpha P^\delta \varepsilon^{*\beta}_\gamma(k)\varepsilon_{Y \mu}(P) , 
\label{ampl2}
\end{equation}
where 
\begin{align}
        \mathcal{M}_1 &= \frac{g_{YD_1D} \ eg_{D_1D^*\gamma} \ g_{G_0DD^*}}{m_{D^*}^2m_{D_1}^2}\int \frac{d^4q}{(2\pi)^4} \ \Pi (P,k,q)\ m_{D_1}^2 (m_{D^*}^2 - m_Y^2  -2q^2 \nonumber
\\ & + 3P\cdot q + 2 P\cdot k - 3 k\cdot q) , \nonumber
\\  \nonumber
\mathcal{M}_2^\delta &=\frac{g_{YD_1D} \ eg_{D_1D^*\gamma} \ g_{G_0DD^*}}{m_{D^*}^2m_{D_1}^2}\int \frac{d^4q}{(2\pi)^4} \ \Pi (P,k,q)\ m_{D_1}^2(- 2m_{D^*}^2 + m_Y^2 + 2q^2 \nonumber
\\
& - 3P\cdot q - 2 P\cdot k + 3 k\cdot q ) q^\delta, \nonumber
\\
\mathcal{M}_3^{\nu \mu} &= \frac{g_{YD_1D} \ eg_{D_1D^*\gamma} \ g_{G_0DD^*}}{m_{D^*}^2m_{D_1}^2} \int \frac{d^4q}{(2\pi)^4}  \ \Pi (P,k,q)\ ( q^\mu q^\nu m_{D^*}^2) , 
\label{ampl3}
    \end{align} 
with 
\begin{equation}
        \Pi(P,k,q) = \frac{1}{[q^2 - m_D^2][(P - q)^2 - m_{D_1}^2][(P - q - k)^2 - m_{D^*}^2]} .
\label{ampl4}
\end{equation}
In the following, we work in the $Y(4230)$ rest frame. After applying the Passarino--Veltman method, the amplitude is given by
\begin{equation}
    \mathcal{M} = (\mathcal{M}_1 + \mathcal{A} + \mathcal{B}) \ \epsilon_{\alpha\beta\nu\delta} \ k^\alpha \varepsilon^{*\beta}_{\gamma}  \varepsilon_Y^\nu P^\delta
\end{equation}
where $\mathcal{M}_1$ is defined in Eq.~\eqref{ampl3}, $\mathcal{A}  = \frac{(k \cdot M_2)}{(k\cdot P)} $ and 
\begin{align}
\mathcal{B} &=  \frac{g_{YD_1D} \ eg_{D_1D^*\gamma} \ g_{G_0DD^*}}{2 m_{D_1}^2}\int \frac{d^4q}{(2\pi)^4} F(|\vec{q}|^2) \ \Pi(P,k,q) 
 \bigg( q^2 - \frac{2( q \cdot k )(q \cdot P ) }{k\cdot P} +  \frac{( q \cdot k )^2 P^2}{(k\cdot P)^2}\bigg) . 
\label{ampl5}
    \end{align} 

The integration over $q^0$ can be performed analytically using Cauchy's theorem. To this end, we make the dependence of the integrands on the $q^0$ variable explicit by expressing the function $\Pi(P,k,q)$ in the form
\begin{align}
\Pi (P,k,q)  & =  
\frac{1}{[q^0 - \omega_D (\vec{q}) + i\epsilon_1][q^0 + \omega_D (\vec{q}) - i\epsilon_1]} 
\nonumber \\ & 
\times \frac{1}{[m_Y - q^0 - \omega_{D_1} (\vec{q}) + i\epsilon_2][m_Y -q^0 + \omega_{D_1} (\vec{q}) - i\epsilon_2]} 
\nonumber \\ & 
\times\frac{1}{[ m_Y -q^0 - k^0 -\omega_{D^*} (\vec{q}+\vec{k}) + i\epsilon_3] [ m_Y -q^0 - k^0 + \omega_{D^*} (\vec{q}+\vec{k}) - i\epsilon_3]} ,  
\label{Pi1}
\end{align}
where $\omega_{\bar D^{(*)}} (\vec{q}) = \sqrt{m_{\bar D^{(*)}}^2  + \vec{q}^2} $ and $ \omega_{D_1}(\vec{q}) = \sqrt{m_{D_1}^2 - im_{D_1}\Gamma_{D_1}+ \vec{q}^2}$. 
Here $\Gamma_{D_1}$ is the total width of the $D_1$ meson, taken as $\Gamma_{D_1} = 31.3$~MeV~\cite{ParticleDataGroup:2024cfk}. This modification ultimately leads to a mild suppression of the decay width.
The poles are therefore: $q^0 = \pm [\omega_D(\vec{q}) -i\epsilon_1]$, $q^0 = m_Y \pm [ \omega_{D_1} (\vec{q}) -i\epsilon_2]$ and $q^0 = m_Y - k^0 \pm [  \omega_{D^*} (\vec{q}+\vec{k}) -i\epsilon_3]$. This integral over $q^0$ is evaluated by choosing the contour in the lower part of the complex $q^0$-plane. 
Thus, after integrating over $q^0$, the amplitude in Eq.~\eqref{ampl5} is given by 

\begin{align}
\mathcal{M} &= \frac{-ig_{YD_1D} \ eg_{D_1D^*\gamma} \ g_{G_0DD^*}}{m^2_{D_1} m^2_{D*}}\epsilon_{\alpha\beta\nu\delta}k^\alpha \varepsilon^{*\beta}_{\gamma}\varepsilon_Y^\nu P^\delta \int \frac{d^3q}{(2\pi)^3} F(|\vec{q}|^2)  \nonumber
    \\
    & \bigg\{ \left[ \mathbb{L}^{(0)} + a_1^{(0)} + b_1^{(0)} \right] \left(\Pi_{\bar{D}}^{(0)}+\Pi_{D_1}^{(0)}+\Pi_{D^*}^{(0)} \right) + \left[ \mathbb{L}^{(1)} + a_1^{(1)} \right]\left(\Pi_{\bar{D}}^{(1)}+\Pi_{D_1}^{(1)}+\Pi_{D^*}^{(1)} \right) \nonumber
    \\
    & + \left[ \mathbb{L}^{(2)} + a_1^{(2)} \right]\left(\Pi_{\bar{D}}^{(2)}+\Pi_{D_1}^{(2)}+\Pi_{D^*}^{(2)} \right) + a_1^{(3)}\left(\Pi_{\bar{D}}^{(3)}+\Pi_{D_1}^{(3)}+\Pi_{D^*}^{(3)} \right) \bigg\} ,    
\label{ampl6}
\end{align}
where
\begin{align}
    \Pi_{\bar{D}}^{(n)} & = \lim_{\epsilon_1,\epsilon_3 \to 0}\frac{(\omega_D)^n}{(2\omega_D)} \frac{1}{(m_Y - \omega_D - \omega_{D_1})(m_Y - \omega_D + \omega_{D_1})}\nonumber
     \\
     &\times\frac{1}{(m_Y - \omega_D - k^0 - \omega_{D^*} +i(\epsilon_3 + \epsilon_1) )(m_Y - \omega_D - k^0 + \omega_{D^*} -i(\epsilon_3 -\epsilon_1))} , \nonumber
\\
    \Pi_{D_1}^{(n)} & = \frac{1}{(2\omega_{D_1})}\frac{( m_Y + \omega_{D_1})^n}{( m_Y + \omega_{D_1} - \omega_D)( m_Y + \omega_{D_1} +\omega_D)}\nonumber 
     \\
     &\times\frac{1}{(- \omega_{D_1} - k^0 - \omega_{D^*} )(- \omega_{D_1}  - k^0 + \omega_{D^*} )} , \nonumber
\\
    \Pi_{D^*}^{(n)} & = \lim_{\epsilon_1,\epsilon_3 \to 0} \frac{1}{( 2\omega_{D^*} )}\frac{(m_Y - k^0 +\omega_{D^*})^n}{(m_Y - k^0 +\omega_{D^*} - \omega_D +i(\epsilon_1 -\epsilon_3))(m_Y - k^0 +\omega_{D^*} +\omega_D -i(\epsilon_1 + \epsilon_3))} \nonumber
     \\
     &\times\frac{1}{( k^0 - \omega_{D^*} - \omega_{D_1})( k^0 -\omega_{D^*} + \omega_{D_1})} , \nonumber
     \\
    \mathbb{L}^{(0)} & = m_{D_1}^2(m_{D^*}^2 - m_Y^2 +2|\vec{q}|^2 +2m_Yk^0 + 3 \vec{k} \cdot \vec{q}, \ \ \ 
    \mathbb{L}^{(1)} = -3m_{D_1}^2 (k^0 - m_Y),   \ \ \ 
    \mathbb{L}^{(2)} = -2m_{D_1}^2, \nonumber
    \\
    a_1^{(3)} &= 2\frac{m_{D_1}^2}{m_Y}, \ \ \
    a_1^{(2)} = 3\frac{m_{D_1}^2}{m_Y}(k^0-m_Y) - 2 \frac{\vec{k} \cdot \vec{q}}{m_Y k^0} m_{D_1}^2, \nonumber
    \\
    a_1^{(1)} &= \frac{m_{D_1}^2}{m_Y}( m_Y^2 -2|\vec{q}|^2 -2m_Y k^0  -3\vec{q}\cdot \vec{k} -2 m^2_{D*}) - 3\frac{\vec{k} \cdot \vec{q}}{m_Y k^0} m_{D_1}^2 ( k^0 - m_Y), \nonumber
    \\
    a_1^{(0)} &= - \frac{\vec{k} \cdot \vec{q}}{m_Y k^0} m_{D_1}^2 (m_Y^2 -2|\vec{q}|^2 -2m_Y k^0  -3\vec{q}\cdot \vec{k} -2 m^2_{D*}), \nonumber
    \\
    b_1^{(0)}& = -\frac{m_{D^*}^2}{2}\bigg[ |\vec{q}|^2 - \frac{(\vec{q} \cdot \vec{k})^2}{{k^0}^2}\bigg]. 
\label{ampl8}
\end{align}

The remaining integration over the three-momentum $\textbf{q}$ is performed including a Gaussian form factor in order to regulate UV divergences in the loop integral and encode the effect of short-distance physics not explicitly included in the model, by means of the prescription $\int   d^3 q \to \int   d^3 q \exp{\left(-\frac{2 |\vec{q}|^2}{\Lambda^2}\right)} $, where $\Lambda$ is a free parameter~\cite{Brandao:2023vyg,Abreu:2025mhl}.


Hence, putting together all the ingredients discussed above, the partial decay width for the $Y(4230) \to \gamma \ G_0(3900)$ reaction is obtained using the expression
\begin{align}
	\Gamma &= \frac{1}{W} \int_{(m_{G_0} - 2 \Gamma_{G_0})^2}^{(m_{G_0} + 2 \Gamma_{G_0})^2}\mathrm{d}s  \ \frac{1}{\pi} \mathrm{Im}\left( \frac{-1}{s-m_{G_0}^2+ i m_{G_0}\Gamma_{G_0}}\right)\nonumber\\
	& \times \frac{|\vec{k}|}{24\pi m_Y^2} \overline{|\mathcal{M}(m_{G_0} \to \sqrt{s})|^2},
\label{gammanum}
\end{align}
where $1/W$ is the factor introduced to normalize the spectral function of the $G_0$, with $W$ defined by
\begin{align}
W = \frac{1}{\pi} \int_{(m_{G_0} - 2 \Gamma_{G_0})^2}^{(m_{G_0} + 2 \Gamma_{G_0})^2}\mathrm{d}s  \  \mathrm{Im}\left( \frac{-1}{s-m_{G_0}^2+ i m_{G_0}\Gamma_{G_0}}\right);
\label{gammanum2}
\end{align}
$\Gamma_{G_0}$ is the $G_0$ decay width, estimated from the  hadronic  $G_0$ width predictions of Ref.~\cite{Liu:2025sjz}, and taken here to be $\Gamma_{G_0} \sim 5 $ MeV; 
$|\vec{k}|$ is the three-momentum of the photon, given by
\begin{eqnarray}
|\vec{k}| = \frac{\sqrt{\lambda(m_Y^2,0,m_{G_0}^2 \to s)}}{2m_Y},
\end{eqnarray}
with $\lambda(a,b,c)$ being the usual Källén function and the mass of $G_0$ replaced by $ \sqrt{s}$; and in the same way $\overline{|\mathcal{M}(m_{G_0} \to \sqrt{s})|^2}$ is the squared modulus of the amplitude, summed over the polarizations of the $Y(4230)$ and the photon.

\section{Results}\label{results}

\begin{figure}[!htbp]
    \centering
    \includegraphics[width=0.8\linewidth]{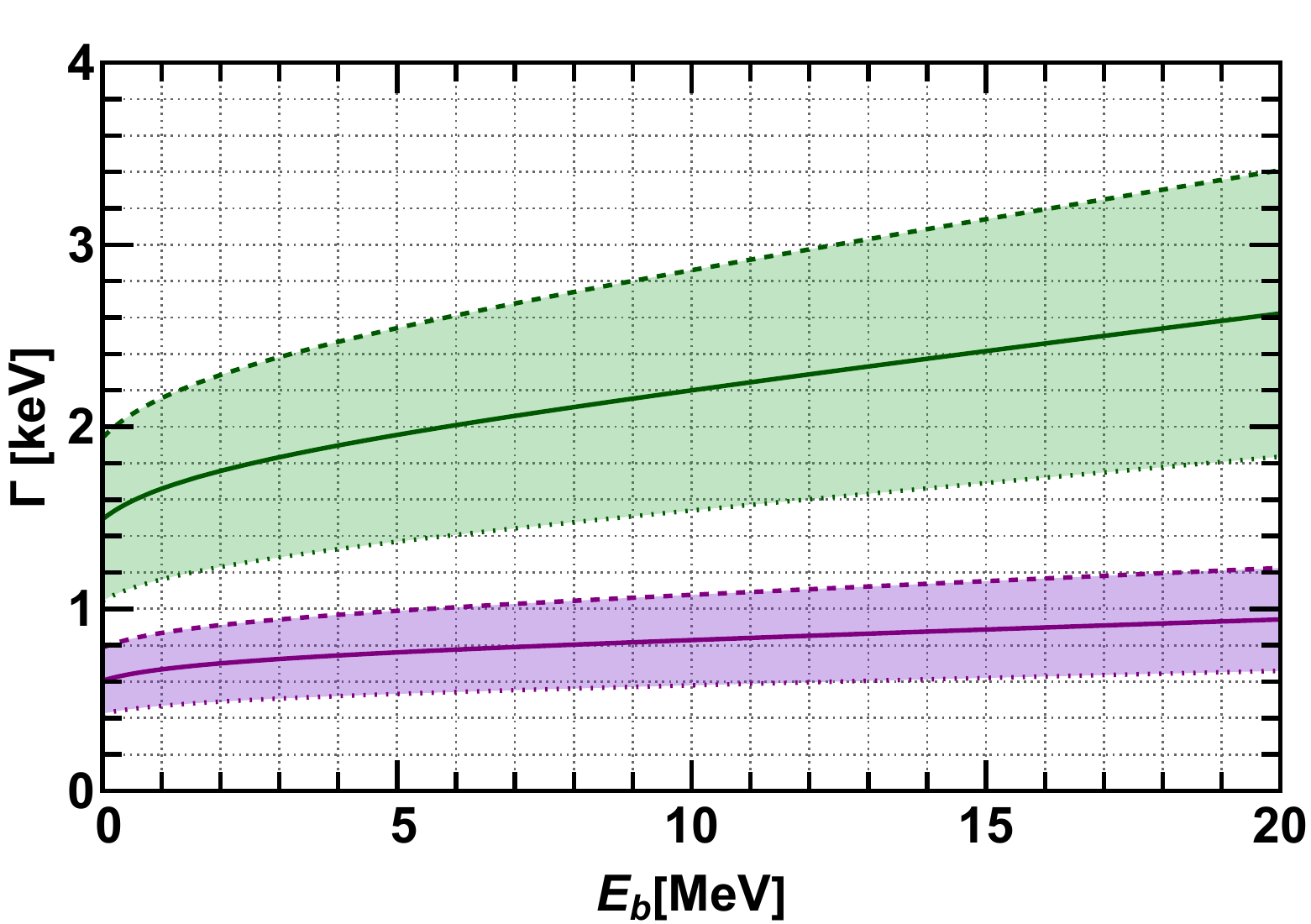}
    \caption{
    The partial decay width $\Gamma$ is shown as a function of the binding energy $E_b$. The dotted, thick, and dashed curves represent the lower bound, central, and upper bound values of the $g_{D_1D^*\gamma}$ coupling, respectively. The green shaded band corresponds to $\Lambda = 1200$~MeV, while the purple shaded band corresponds to $\Lambda = 800$~MeV. $\Lambda_g$ is fixed at $1000$~MeV.}
    \label{fig:bind_band}
\end{figure}

\begin{figure}[!htbp]
    \centering
    \includegraphics[width=0.8\linewidth]{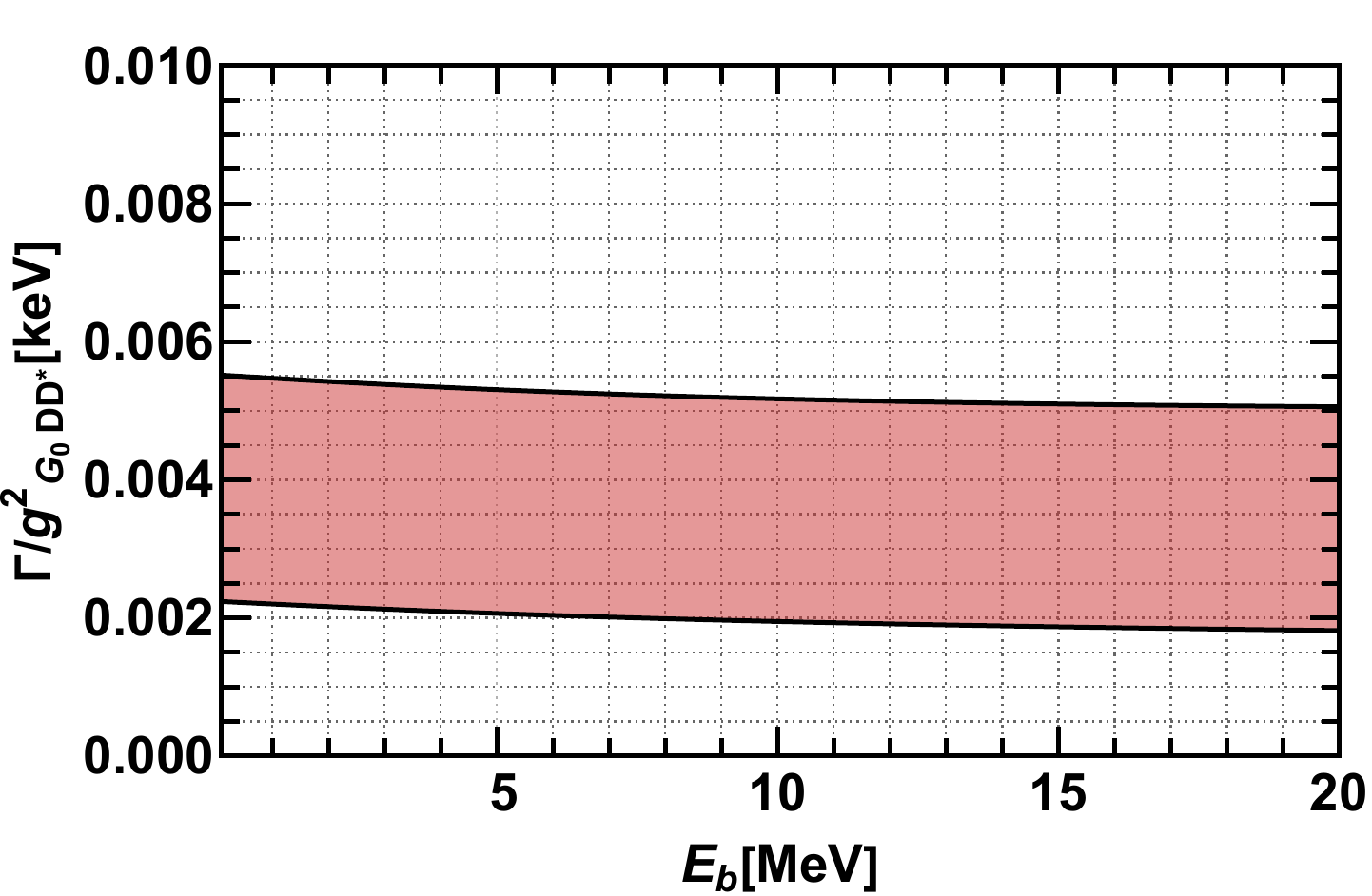}
    \caption{Reduced partial decay width $\Gamma / g_{G_0DD^*}^2$ as a function of $E_b$. The upper and lower bold black curves correspond to $\Lambda = 1200$ MeV and $\Lambda = 800$ MeV, respectively, with  $g_{D_1D^*\gamma}$ set to its central value. }
    \label{fig:GammaOverg2squared x Eb}
\end{figure}

\begin{figure}[!htbp]
    \centering
    \includegraphics[width=0.8\linewidth]{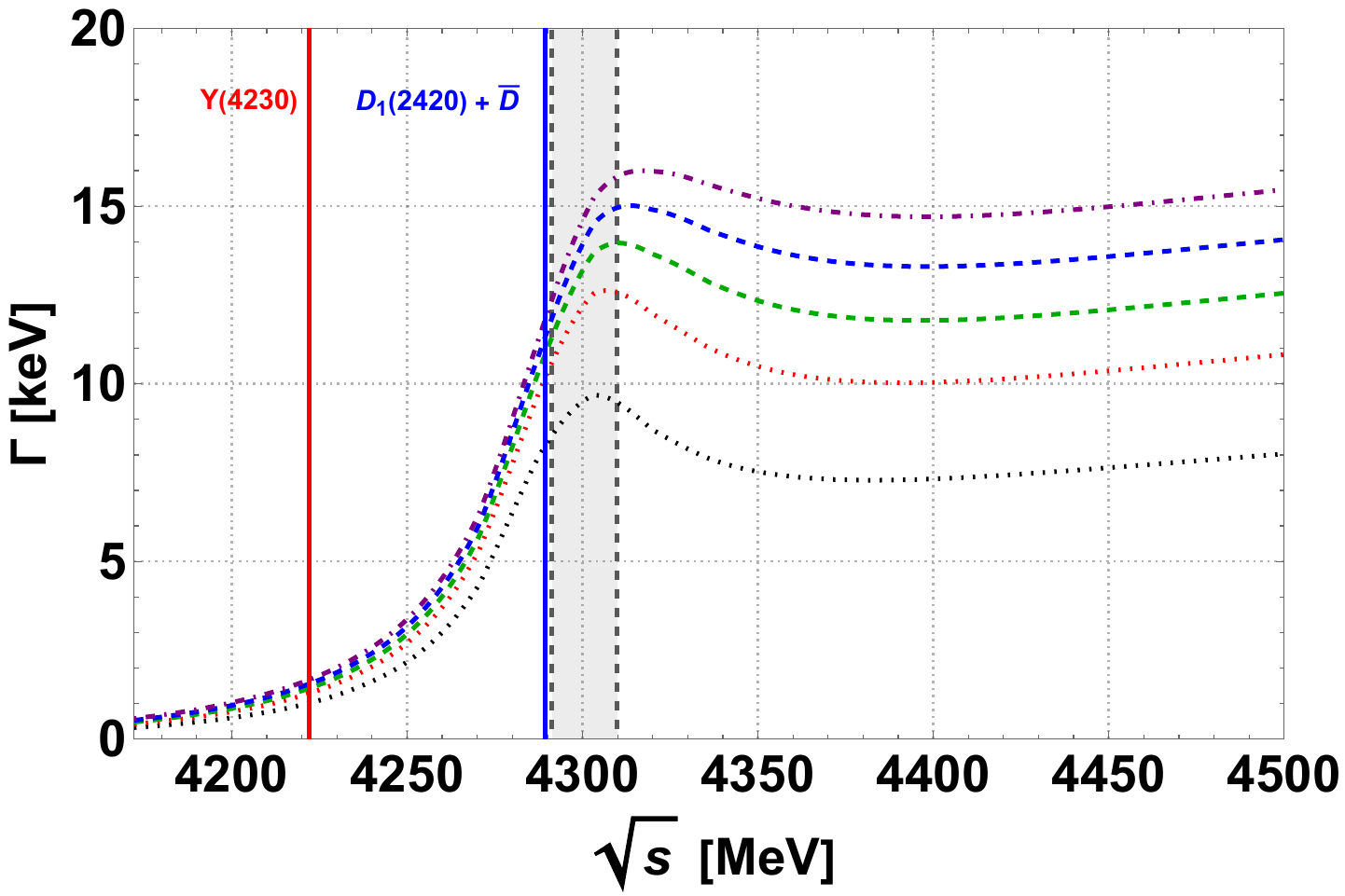}
    \caption{Partial decay width $\Gamma$ as a function of the invariant mass of the system $\gamma G_0$, for different values of the binding energy of $G_0$. The curves have been calculated with $\Lambda = 1000$~MeV, $\Lambda_g = 1000$ MeV and $g_{D_1D^*\gamma}$ set to its central value. The dotted black, dotted red, dashed green, dashed blue and dot-dashed purple curves correspond to the results for $E_b = 0.1, 5, 10, 15 $ and 20 MeV. 
    The red and blue vertical lines denote the $Y(4230)$ mass and the threshold of the $D_1\bar{D}$ system, respectively, while the shaded area indicates the region where a triangular singularity may occur.
    }
    \label{fig:mY}
\end{figure}

In this section, we present the results for the partial decay width of $Y(4230) \to \gamma G_0(3900)$ given in Eq.~\eqref{gammanum} and its dependence on the model parameters. Since $G_0(3900)$ has not yet been observed, its mass remains experimentally unknown. We therefore parametrize it as $m_{G_0} = m_{D^*} + m_D - E_b$, where $E_b$ is the binding energy, and explore a range of shallowly bound molecular configurations with $E_b$ between $0.1$ and $20$~MeV. 
The numerical values for the isospin-averaged meson masses are taken from the Review of Particle Physics (RPP)~\cite{ParticleDataGroup:2024cfk}. 
The size parameter $\Lambda$ is chosen within the interval usually employed in other similar effective approaches:  $\Lambda \sim 0.8 - 1.2 \ \mathrm{GeV}$~\cite{Liu:2024ziu,Wu:2023rrp,Abreu:2025mhl,Brandao:2025cli}. 
Concerning the other size parameter, $\Lambda_g$, we adopt the central value $\Lambda_g = 1.0$~GeV for the estimates shown in Figs.~\ref{fig:bind_band}-\ref{fig:mY}; while in the findings presented in Table~\ref{Tablecoupl-eb} we employ the range $\Lambda_g \sim 0.8 - 1.2 \ \mathrm{GeV}$, aligned with that of Ref.~\cite{Liu:2025sjz}. We refer the reader to Appendix~\ref{appa} for a discussion of the different roles of $\Lambda$ and $\Lambda_g$. 

Furthermore, as noted in Ref.~\cite{Liu:2024ziu}, the coupling $g_{D_1D^*\gamma}$ that appears in the vertices of the triangle diagrams is subject to substantial theoretical uncertainty, owing to its extraction from model-dependent estimates of the $D_1$ radiative widths. This uncertainty, estimated to be of order $30\%$, has been propagated throughout our numerical analysis.



\begin{figure}[!htbp]
    \centering
    \includegraphics[width=0.8\linewidth]{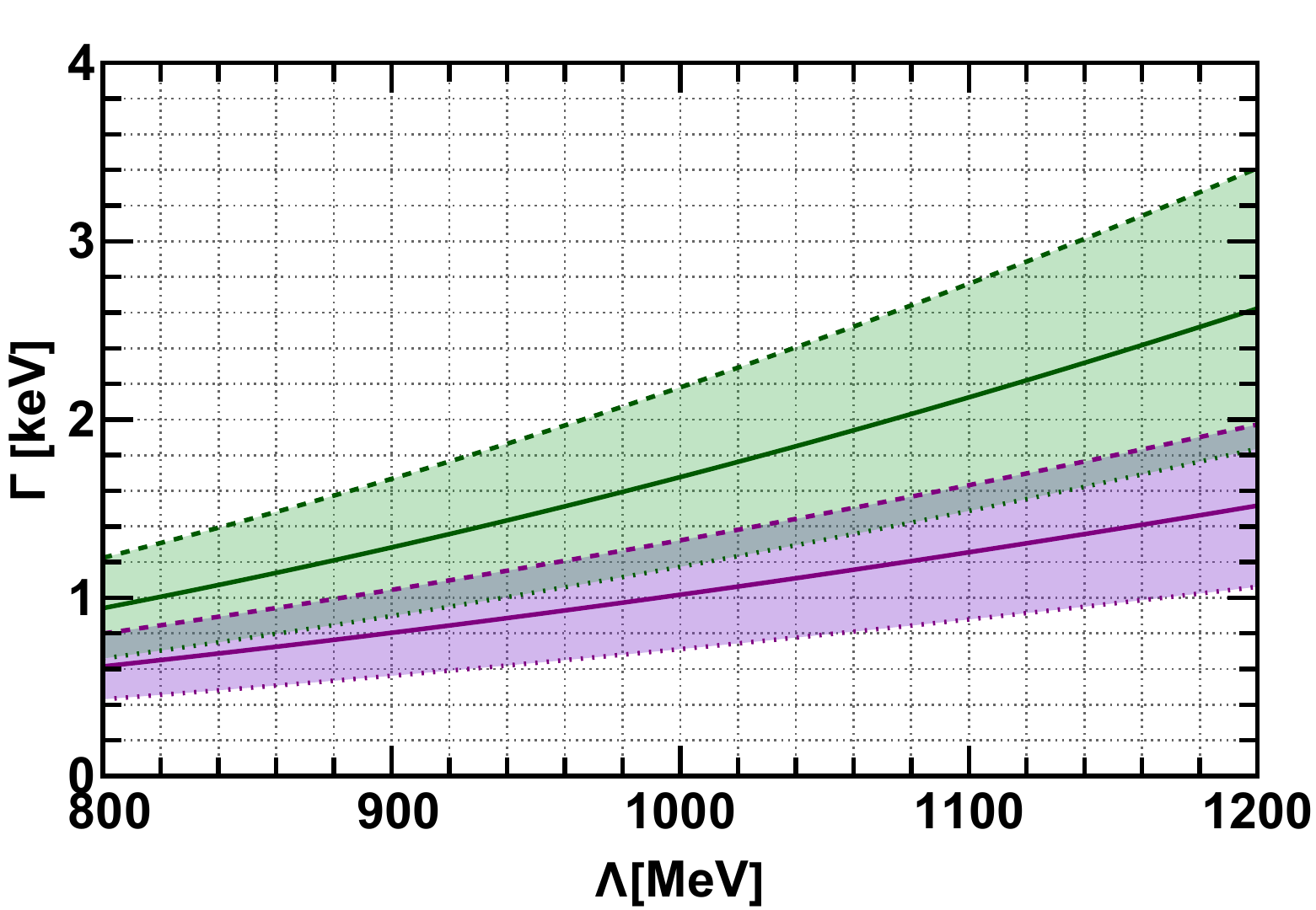}
    \caption{
    Partial decay width $\Gamma$ as a function of $\Lambda$. The dotted, thick, and dashed curves correspond to the lower bound, central, and upper bound values of the $g_{D_1D^*\gamma}$ coupling, respectively. The green and purple shaded bands correspond to $E_b = 20$~MeV and $E_b = 0.1$~MeV, respectively. The cutoff $\Lambda_g$ is fixed at $1000$~MeV.}
    \label{fig:lambda_band}
\end{figure}

Fig.~\ref{fig:lambda_band} shows the partial decay width $\Gamma$ as a function of $\Lambda$ for fixed values of $E_b$. The green and purple shaded bands correspond to $E_b = 20$~MeV and $E_b = 0.1$~MeV, respectively. As expected, $\Gamma$ increases with $\Lambda$. Quantitatively, varying $\Lambda$ from its lower to its upper bound changes the value of $\Gamma$ by roughly a factor of $2.6$. More concretely, depending on the choice of $E_b$, the resulting partial width $\Gamma$ ranges from \Pedro{tenths to units of keV over the considered $\Lambda$ interval}.

Fig.~\ref{fig:bind_band} shows the partial width $\Gamma$ as a function of the binding energy $E_b$. The results reveal that $\Gamma$ exhibits a monotonic increase up to the upper limit of $E_b$ considered in our analysis. 
To clarify the role of phase-space versus coupling effects in the $E_b$ dependence of the radiative width, we separate the effects on the coupling constant $g_{G_0DD^*}$ from those on the reduced partial decay width $\Gamma / g_{G_0DD^*}^2$. The coupling $g_{G_0DD^*}(E_b)$ is determined in Appendix~\ref{appa}, and from Fig.~\ref{fig:gDDstarG0 x Lambda g} one can infer that it increases with $E_b$. The behavior of $\Gamma / g_{G_0DD^*}^2$ as a function of $E_b$ is shown in Fig.~\ref{fig:GammaOverg2squared x Eb}. As can be seen, the reduced decay width exhibits only a mild decrease with increasing $E_b$, confirming that the growth of the radiative width with $E_b$ is primarily driven by the behavior of $g_{G_0DD^*}$.

Besides, we have also studied the dependence of the radiative width $\Gamma[Y(4230) \to \gamma G_0]$ on the assumed $G_0$ width, varying $\Gamma_{G_0}$ in the range 0.1--5 MeV. Our findings indicate no significant sensitivity to $\Gamma_{G_0}$. This insensitivity can be attributed to the small width relative to the typical energy scales, so the convolution of the squared amplitude with the Breit-Wigner distribution does not produce a noticeable variation. 

We also compare these findings with other studies of exotic hadron production via $Y(4230)$-radiative transitions in the literature. 
As discussed in the Introduction, Refs.~\cite{Chen:2025gxe,Lin:2024qcq,Liu:2025sjz} have explored different interpretations for the $G_0(3900)$. Nevertheless, all of them are compatible with the bound-state scenario assumed in the present work, albeit each study examined different ranges of binding energy. Here we adopt the range $E_b = 0.1$--$20$~MeV below the $D\bar{D}^*/\bar{D}D^*$ threshold and a size parameter $\Lambda \in [800,1200]$~MeV.
Accordingly, using the central value of the $Y(4230)$-decay width from Ref.~\cite{ParticleDataGroup:2024cfk}, $\Gamma_{Y(4230)} = 51 $ MeV, the decay branching fraction for this process is  
\begin{eqnarray}
 \mathcal{B} =  0.98 \times 10^{-5} - 7.7 \times  10^{-5}. 
    \label{br_fr}
\end{eqnarray}
In Table~\ref{Tablecoupl-eb}, we explicitly show the predicted values of the decay width and the branching ratio for specific binding energies. 

\begin{table}[!htbp] 
\centering
\caption{
Predicted values of the decay width $  \Gamma $ and the branching ratio $ \mathcal{B}$ of the reaction $Y(4230) \to \gamma  G_0(3900)$, for some values of the binding energy $E_b$. Values correspond to $\Lambda = 0.8  - 1.2$~GeV on the left - right. }
\begin{tabular}{c| c | c | c | c}
\hline
\hline
$E_b$ [MeV]   &  $ \Gamma_{\Lambda_g = 800}$  [keV] &  $ \Gamma_{\Lambda_g = 1200}$  [keV] & $\mathcal{B}_{\Lambda_g = 800}$ $( \times 10^{-5})$ & $\mathcal{B}_{\Lambda_g = 1200}$ $( \times 10^{-5})$ \\ \hline
0.1 & 0.79  $-$ 1.95 & 0.50 $-$ 1.24 & 1.55  $-$ 3.83  & 0.98 $-$ 2.42 \\
1   & 0.88 $-$ 2.18  & 0.54 $-$ 1.33 & 1.72 $-$ 4.27  & 1.05  $-$ 2.61 \\
5   & 1.05 $-$ 2.69  & 0.59 $-$ 1.52 & 2.05 $-$ 5.28 & 1.16  $-$ 2.98 \\ 
10  & 1.18 $-$ 3.13  & 0.63 $-$ 1.67 & 2.31  $-$ 6.14  & 1.23  $-$ 3.27 \\
15  & 1.29 $-$ 3.53  & 0.66 $-$ 1.80 & 2.54  $-$ 6.91  & 1.29  $-$ 3.53\\
20  & 1.40 $-$ 3.91  & 0.69 $-$ 1.93 & 2.75  $-$ 7.66  & 1.36 $-$ 3.78\\
\hline\hline
\end{tabular}
\label{Tablecoupl-eb}
\end{table}

It is interesting to note that Ref.~\cite{Liu:2024ziu} studied the decay $Y(4230) \to \gamma X(3872)$, assuming the $X(3872)$ state to be a shallow $S$-wave $\bar{D}D^* + \text{c.c.}$ bound state. Its partial width was estimated to be $2$--$34$~keV, which is of the same order as our result for the \Pedro{largest} binding energy in Table~\ref{Tablecoupl-eb}. It should be stressed, however, that other independent calculations present different values for $\Gamma_{Y(4230) \to \gamma X(3872)}$. For instance, Ref.~\cite{Dong:2014zka} treated both $Y(4230)$ and $X(3872)$ as composite states containing hadronic molecular and compact charmonium components, obtaining $\Gamma_{Y(4230) \to \gamma X(3872)} \approx 23.2$--$48.6$~keV under various mixing scenarios. Adopting a different perspective based on the compact tetraquark picture, Ref.~\cite{PhysRevD.103.094024} evaluated the same decay width to be $\sim 100$~keV. 
As can be seen, these distinct works report partial widths for $X(3872)$ production that are comparable or one order larger than our predictions for the reaction $Y(4230) \to \gamma G_0(3900)$, when considering $G_0(3900)$  with \Pedro{larger} binding energies in Table~\ref{Tablecoupl-eb}.

At this point, it is also worth discussing the relevance (or lack thereof) of triangle singularities in the amplitude studied above. This is an interesting point because some new ``hadron'' candidates do not necessarily reflect the existence of a genuine particle; instead, they may have a dynamical explanation based on how known hadrons rescatter under the strong force. One such mechanism is the triangle singularity, which occurs when three intermediate particles go on-shell, with two of them having parallel momenta, provided the kinematics allows it --- an effect known from the Coleman--Norton theorem. We refer the reader to Refs.~\cite{Landau:1959fi,Coleman:1965xm,Guo:2019twa,Abreu:2020jsl,Llanes-Estrada:2021ath,Abreu:2021xpz,Huang:2025rvj} for a detailed explanation of this mechanism.

In the present case, according to the Coleman--Norton theorem~\cite{Guo:2019twa}, a triangular singularity can only occur if the initial decaying particle in Figs.~\ref{fig:neutral_triangle} and~\ref{fig:charged_triangle} has a mass within the range $4291.3$--$4309.9$~MeV; similarly, the final massive particle must have a mass within $3875.7$--$3892.9$~MeV. The mass of the $Y(4230)$ state used here is $4222.2$~MeV, and the $G_0$ mass lies below the $D\bar{D}^*/\bar{D}D^*$ threshold --- i.e., below the lower limit for the final massive particle. Thus, although the numbers are close, our system does not strictly satisfy the conditions required to produce a triangle singularity. For completeness, Fig.~\ref{fig:mY} shows the partial decay width as a function of the $\gamma G_0$ invariant mass for different values of the binding energy. As can be seen, the prominent peaks in the $\gamma G_0$ production line shapes are located far from the $Y(4230)$ mass. We also note that the peak associated with the triangle singularity is not very prominent, as we have included the width of the intermediate $D_1$ meson in the calculations. Therefore, if the $G_0(3900)$ is experimentally detected in the proposed decay, this finding might be interpreted as evidence that it is a genuine hadron, rather than a kinematical effect, although it does not exclude all
possible threshold or rescattering effects.

We should also note that if the initial decaying particle were, for example, the $Y(4320)$ state recently observed by BESIII with a mass of $4298$~MeV~\cite{BESIII:2022qal}, and if it had a nonnegligible branching ratio for decay into the $D_1 \bar D$ system, it could produce a triangle singularity in the region of interest. This could serve as a complementary analysis to the one proposed here.

Some remarks concerning the nature of the $G_0(3900)$ state are in order. As shown in previous studies~\cite{Lin:2024qcq, Chen:2025gxe}, the $P$-wave $D\bar{D}^*/\bar{D}D^*$ system admits both bound-state and resonant solutions depending on the model parameters and the cutoff. In this work, we adopt the bound-state realization. The extension to the resonant case is beyond the present scope, as it would require a treatment involving a complex propagator and spectral density. Nonetheless, for a narrow resonance close to threshold, the coupling $g_{G_0DD^*}$ acquires a small imaginary part associated with the imaginary part of the pole's position, while the real part remains close to the bound-state value~\cite{ParticleDataGroup:2024cfk}. The resulting corrections to the radiative rates are expected to be small.
Therefore, assuming a smooth continuation from the bound-state to the resonant regime near threshold, we expect the production rates for a resonance to remain of the same order of magnitude as those obtained for the bound state. This expectation relies on the smoothness of the continuation rather than on a full treatment of the resonant case, which lies beyond the scope of this study.
 
\begin{table}[!htbp]
\centering
\caption{
Predicted values of the decay widths $\Gamma$ and the branching ratio $\mathcal{B}$ of the reaction $Y(4230) \to \gamma G_0(3900)$, for some values of the binding energy $E_b$. Values correspond to $\Lambda = 1.0$~GeV. The first values correspond to the charged + neutral loop contributions using physical masses and the couplings $g_{D_1^0 D^{*0}\gamma} + g_{D_1^+ D^{*+}\gamma}$, while the values in parentheses are obtained under the isospin-limit approximation with $g_{D_1D^*\gamma}$ set to its central value. Column $\Delta\mathcal{B}=\frac{|\mathcal{B}{\rm phys}-\mathcal{B}{\rm iso}|}{\mathcal{B}_{\rm phys}}$ shows the relative percentage difference between the two sets of branching ratio values.}
\begin{tabular}{c | c | c | c}
\hline
\hline
$E_b$ [MeV] & $\Gamma_{\Lambda_g=1000}$ [keV] & $\mathcal{B}$ & $\Delta \mathcal{B}$ [\%] \\
\hline
1   & 1.39 (1.11) & $2.71 \times 10^{-5}$ ($2.17 \times 10^{-5}$) & 20.14 \\
5   & 1.47 (1.29) & $2.88 \times 10^{-5}$ ($2.53 \times 10^{-5}$) & 12.24 \\
10  & 1.62 (1.43) & $3.18 \times 10^{-5}$ ($2.81 \times 10^{-5}$) & 11.72 \\
15  & 1.74 (1.56) & $3.40 \times 10^{-5}$ ($3.05 \times 10^{-5}$) & 10.34 \\
20  & 1.86 (1.68) & $3.64 \times 10^{-5}$ ($3.28 \times 10^{-5}$) & 09.67 \\
\hline
\hline
\end{tabular}
\label{Tablecoupl-eb2}
\end{table}

Another point deserving a comment is the adoption of the isospin limit. This is a reasonable simplification for this first exploratory calculation, especially given that no experimental searches for this reaction have been reported in the literature to date, and the present work is intended to serve as a guide for future experimental investigations. As this is the first theoretical estimate for the radiative width of the $G_0$ production, we believe that such simplifications are appropriate at this stage.

Nevertheless, to roughly assess the impact of isospin-breaking effects, we have evaluated the neutral and charged loop contributions separately, using physical masses and couplings for a few representative parameter choices, and compared them with the isospin-limit result. The results are shown in Table~\ref{Tablecoupl-eb2}.
As can be observed, the resulting branching ratios are very close for larger values of the binding energy, i.e., farther from the $D\bar{D}^{*}/D^{*}\bar{D}$ thresholds. However, for smaller values of $E_b$, the neutral ($D^0\bar{D}^{*0}/D^{*0}\bar{D}^{0}$) and charged ($D^+D^{*-}/D^{*+}D^{-}$) thresholds differ, and the couplings $g_{D_1^0 D^{*0}\gamma}$ and $g_{D_1^+ D^{*+}\gamma}$ also deviate. As a consequence, these effects have a noticeable impact, leading to larger uncertainties in the branching ratio at low binding energies. 
We should notice that this difference should be identified as the change within this specified test, rather than a general estimate of all isospin-breaking effects on the pole and its couplings.
This uncertainty, nevertheless, is comparable to or smaller than other theoretical uncertainties in our calculation, such as those stemming from $E_b$, $\Lambda_g$, and $g_{D_1D^*\gamma}$. We therefore consider the isospin limit to be a reliable baseline for this first estimate, leaving a full coupled-channel treatment for future work.

Ultimately, the small branching fraction obtained in Eq.~\eqref{br_fr} renders the observation of this radiative transition a challenging task for experimental collaborations, to be addressed in the future. Nevertheless, we hope that this result, together with those mentioned above, will help guide experimental facilities in future searches for the $G_0(3900)$ and in testing its structure as a $P$-wave molecular state.





\section{Conclusions}
\label{Conclusions}

In this work, we have investigated the radiative decay $Y(4230) \to \gamma G_0(3900)$ as a complementary probe of the $G_0(3900)$ state. The $G_0(3900)$ has been interpreted as a $P$-wave molecular state with quantum numbers $J^{PC}=0^{-+}$, dominated by the $D\bar{D}^{\ast}/\bar{D}D^{\ast}$ components. We have considered the triangle diagram mechanism in which $Y(4230)$ decays into $D_1\bar{D}$, followed by the radiative transition $D_1 \to D^*\gamma$ and the rescattering of $D^*\bar{D}$ into the $G_0(3900)$. Using an effective Lagrangian framework evaluated in the isospin limit, we have computed the partial decay width for this process, focusing on the sensitivity to two key parameters: the binding energy $E_b$ of the $G_0(3900)$ and the parameter $\Lambda$ of the form factor.

Our results show that the partial width $\Gamma(Y(4230) \to \gamma G_0(3900))$ ranges from tenths to units of keV, depending on the choice of relevant quantities ($E_b$, $\Lambda$, $\Lambda_g$ and the coupling uncertainties). 
The width increases monotonically with $\Lambda$ by a factor of approximately $2.6$ over its allowed range, whereas it shows a milder dependence on the binding energy, with a slower growth.
Comparisons with existing literature on the analogous decay $Y(4230) \to \gamma X(3872)$ reveal that our predicted branching fraction for $G_0(3900)$ as a shallow bound state is of the same order of magnitude relative to that of $X(3872)$, depending on the assumed internal structure of the states and approach used. 

The small branching fraction obtained in this study makes the radiative transition $Y(4230) \to \gamma G_0(3900)$ a challenging channel to observe at current experimental facilities,
especially due to the presence of some dominant background processes, such as $Y(4230) \to \gamma X(3872)$ and $Y(4230) \to \gamma \eta_c (1S/2S)$.
Furthermore, on the theoretical side, another point deserving further discussion is that the approach above considered only the $(D\bar{D}^* - \text{c.c.})$ configuration,  leaving other possible components---such as $(D_s\bar{D}_s^* - \text{c.c.})$ or even a compact $c\bar{c}$ core~\cite{Dong:2009uf,Dong:2009yp}---unaccounted for. 
In principle, these other channels would be relevant. However, since this state has not yet been observed, we view the present work as a first step toward a rough estimate of the contribution from the presumably dominant channels.

Future studies incorporating additional components will be important to refine the present estimates. 

Nevertheless, our results provide a concrete, parameter-dependent estimate for this decay channel and highlight the sensitivity of the partial width to the molecular binding scenario. We hope that this work, together with complementary theoretical and experimental efforts, will help guide future searches for the $G_0(3900)$ and contribute to a deeper understanding of its nature as a $P$-wave $D\bar{D}^{*}$ molecular candidate. In particular, improved lattice QCD calculations or future data from BESIII, Belle II, and LHCb could help constrain the binding energy and the size parameter, thereby sharpening the predictions presented here.

\section{Acknowledgement}
L.S.G. acknowledges support from the National Science Foundation of China under Grants Nos.~W2543006 and 12435007. Y.W.P acknowledges support from the National Natural Science Foundation of China under Grant No. 12547153 and the China Postdoctoral Science Foundation under Grant Number 2025M784262 and GZC20261844.

\appendix
\section{Calculation of the coupling $g_{G_0DD^{*}}$}
\label{appa}

The coupling constant $g_{G_{0}DD^{*}}$ is determined via Weinberg's compositeness criterion~\cite{Weinberg:1965zz}. To this end, we follow the procedure used in Ref.~\cite{Liu:2025sjz} and employ the non-local version of the effective Lagrangian in Eq.~\eqref{Lagr1}, 
\begin{align}
\mathcal{L}_{G_{0}DD^{*}} &= \frac{1}{\sqrt{2}} g_{G_{0}DD^{*}} G_{0}(x) \int d^{4}y\, \Phi(y^{2}) \nonumber \\
&\times \Big[ D(x+\omega_{D^{*}}y)\,\overset{\leftrightarrow}{ \partial}{}^{ \mu}\, \bar{D}^{*}_{\mu}(x-\omega_{D}y) \nonumber \\
&\quad - \bar{D}(x+\omega_{D^{*}}y)\, \overset{\leftrightarrow}{\partial}{}^{ \mu}\, D^{*}_{\mu}(x-\omega_{D}y) \Big],
\label{Lag_G0}
\end{align}
where $y$ is the relative Jacobi coordinate, $\omega_{i} = m_{i}/(m_{i}+m_{j})$, and $\Phi(y^{2})$ is a correlation function that describes the finite spatial extension of the molecular state. Its Fourier transform $\tilde{\Phi}(q_{E}^{2})$ is chosen as the Gaussian form $\tilde{\Phi}(q_{E}^{2}) = \exp(-q_{E}^{2}/\Lambda_g^{2})$, with $\Lambda_g$ being a cutoff.

Here is important to distinguish between the two cutoff parameters that appear in our calculation, namely $\Lambda_g$ and $\Lambda$. The cutoff $\Lambda_g$ enters the nonlocal $G_0 D \bar{D}^*$ vertex through the Gaussian correlation function in Eq.~\eqref{Lag_G0} and controls the short-distance behavior of the effective coupling $g_{G_0 DD^*}$ extracted from the compositeness criterion. Physically, $\Lambda_g$ is a parameter of the bound-state wave function (or, equivalently, of the effective theory used to describe the molecular state), and it sets the scale at which the internal structure of the molecule is resolved. In contrast, the cutoff $\Lambda$ appears in the Gaussian form factor employed in the production loop, introduced to control the ultraviolet behavior of the production amplitude and is therefore a parameter of the loop integral rather than of the bound-state wave function. In principle, $\Lambda_g$ and $\Lambda$ need not be identical, as they parametrize different physical scales: the size of the molecular state versus the short-distance dynamics of the triangle loop. In the absence of a first-principles derivation of either, and in the spirit of a phenomenological estimate, we treat them as independent parameters and study the sensitivity of our results to each. The assumption underlying this treatment is that both cutoffs are of the same order, reflecting the typical hadronic scale, and are chosen in the usual range employed in other previous effective approaches (see for example discussions in Refs.~\cite{Liu:2024ziu,Wu:2023rrp,Abreu:2025mhl,Brandao:2025cli}).

According to Weinberg's compositeness criterion~\cite{Weinberg:1965zz},  the coupling constant $g_{G_{0}DD^{*}}$  is fixed by requiring the wavefunction renormalization constant of the composite state $G_{0}$ to vanish: 
\begin{equation}
Z = 1 - \left.\frac{d\Sigma}{dp^{2}}\right|_{p^{2}=m_{G_{0}}^{2}} = 0,
\label{WBcomp1_G0}
\end{equation}
where $\Sigma$ is the mass operator (self-energy) of the $G_{0}$, given by
\begin{align}
     \Sigma &= - g_{G_{0}DD^{*}}^{2} \int \frac{d^4q}{(2\pi)^4} \tilde{\Phi}^2[(q-\omega_Dp)^2]\frac{1}{q^2-m_{D}^2}\frac{\bigg(-g_{\mu\nu} + \frac{(p-q)_\mu (p-q)_\nu}{m_{D^*}^2}\bigg)}{(p-q)^2-m_{D^*}^2}(p-2q)^\mu (p-2q)^\nu .
\label{massop_G0}     
\end{align}

\begin{figure}[!htb]
    \centering
    \includegraphics[width=0.5\linewidth]{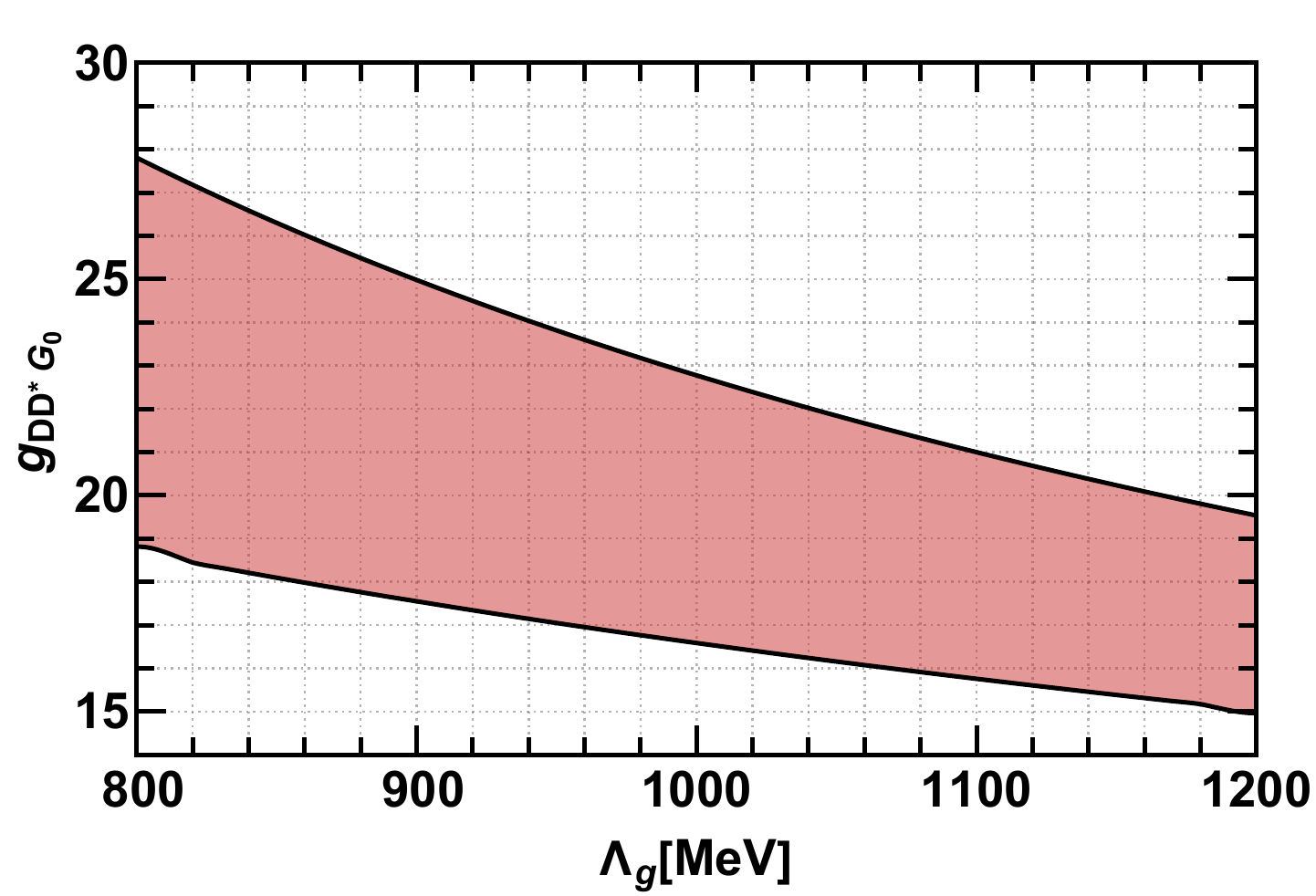}
    \caption{Coupling for the vertex ${DD^*G_0}$ as a function of the coupling cutoff $\Lambda_g$, for different values of the binding energy of $G_0$. The upper and lower black bold lines correspond, respectively, to $E_b = 20$ MeV and $0.1$ MeV. }
    \label{fig:gDDstarG0 x Lambda g}
\end{figure}

Upon performing the loop integral using Schwinger parametrization and applying the condition in Eq.~(\ref{WBcomp1_G0}), the coupling $g_{G_{0}DD^{*}}$ is obtained as a function of $\Lambda_g$ and the binding energy $E_{b}$. The numerical results, displayed in Fig. \ref{fig:gDDstarG0 x Lambda g}, exhibit a monotonic decrease with  $\Lambda_g$ and increasing with  $E_{b}$, in line with those reported in Ref.~\cite{Liu:2025sjz}. These values serve to estimate the theoretical uncertainties in our calculations.

\bibliographystyle{apsrev4-2}
\bibliography{references}

\end{document}